\documentclass[twocolumn]{pazhb_eng}

\usepackage{graphicx}
\usepackage{amsmath}
\usepackage{epsfig,graphics}
\usepackage{rotating}
\usepackage{epsfig}

\newcommand {\be}{\begin{equation}}
\newcommand {\ee}{\end{equation}}

\begin{document}
\journalinfo{2015}{41}{10}{562}[574]

\title{Determination of parameters of long-term variability of the X-ray pulsar LMC X-4}
\author{\bf 
Molkov S.V.\email{molkov@iki.rssi.ru}\address{1}, Lutovinov A.A.\address{1}, Falanga M.\address{2}
\addresstext{1}{Space Research Institute, Moscow, Russia}
\addresstext{2}{International Space Science Institute, Bern, Switzerlandn}
}

\shortauthor{}
\shorttitle{}
\submitted{22 May 2015}

\begin{abstract}
Abstract -- We have investigated the temporal variability of the X-ray flux
measured from the high-mass Xray
binary LMCX-4 on time scales from several tens of days to tens of years, i.e.,
exceeding considerably the orbital period (1.408 days). In particular, we have
investigated the 30-day cycle of modulation of the X-ray
emission from the source (superorbital or precessional variability)
and refined the orbital period and its first derivative. We show that the
precession period in the time interval 1991--2015 is near its equilibrium value
$P_{sup} = 30.370$ days, while the observed historical changes in the phase of this
variability can be interpreted in terms of the ``red noise'' model. We have
obtained an analytical law from which the precession phase can be determined to
within 5\% in the entire time interval under consideration. Using archival data from
several astrophysical observatories, we have found 43 X-ray eclipses in LMC X-4 that,
together with the nine eclipses mentioned previously in the literature, have allowed
the parameters of the model describing the evolution of the orbital period to be
determined. As a result, the rate of change in the orbital period
$\dot P_{orb}/P_{orb}=(1.21\pm0.07)\times10^{-6}$ yr$^{-1}$ has been shown to be
higher than has been expected previously.

\medskip
\end{abstract}

\section*{Introduction}

The X-ray binary system LMC X-4 is located in the Large
Magellanic Cloud (LMC), the nearest galaxy
to us (the distance to it is $\sim50$ kpc), and was discovered
in X rays by the UHURU space observatory (Giacconi
et al. 1972). Subsequently, a 14th magnitude
optical OB star located in the source's X-ray error
circle was proposed as the normal companion to the
relativistic object (Sanduleak and Philip 1977). Optical
photometric and spectroscopic studies of this star
(Chevalier and Ilovaisky 1977; Hutchings et al. 1978)
conclusively proved that the system is a binary, and
the orbital period $P_{orb}\simeq1.408$ days was determined
from the modulation of its optical emission. This
result was also confirmed in X-rays
based on the observation of eclipses (Li et al. 1978;
White 1978), suggesting a high binary inclination.
Observations show that many X-ray sources exhibit
a long term-variability in their emission on time
scales from several tens of days to several years.
However, LMC X-4 is one of the few sources whose
long-term variability has a distinct periodicity. The
X-ray pulsars Her X-1 and SMC X-1 and the microquasar
SS 433 are other objects exhibiting a similar
behavior. Variations in the flux registered from
LMC X-4 with a period $P_{sup}\simeq30.5$ days (the variability
was named ``superorbital'' in the literature;
below in the text, we will adhere to the name ``precessional''
or 30-day modulation) were first detected in
the data from the instruments of the HEAO1 observatory
(Lang et al. 1981). Several models, with accretion
disk precession and radiation-induced accretion
disk warping (see, e.g., Kotze and Charles 2012,
for a brief review) being the main ones, have been
proposed as an explanation of the 30-day modulation.
Apart from the long-term periodic variations in the X-ray
emission from the binary, aperiodic intense series
of short flares (each with a duration of several tens of
seconds) lasting for about an hour and with an occurrence
frequency of about once in several days are
also observed (see, e.g., Epstein et al. 1977; Levine
et al. 2000; Moon and Eikenberry 2001). During
one of such series of flaring activity a coherent
pulsating emission with a period $P_{spin}\simeq13.5$ s was
detected from the source for the first time (Kelley
et al. 1983). Subsequently, pulsations with this period
arising from the spin of the compact object were also
detected during the ``quiescent'' state; as a result,
LMC X-4 was classified as an X-ray pulsar.

Based on GINGA data, Levine et al. (1991)
showed that the orbital period in LMCX-4 decreases.
Subsequent observations with other instruments and
observatories confirmed this conclusion and allowed
the rate of such a change to be determined
(Safi-Harb et al. 1996; Woo et al. 1996; Levine et al. 2000;
Falanga et al. 2015).

In this paper, we analyzed all the available observational
data for LMC X-4 from the BATSE, RXTE,
MAXI, SWIFT, INTEGRAL, and XMM-Newton
space observatories and instruments and obtained in
total a quasi-continuous time series with a duration
of around 25 years (1991--2015). Using these
data and historical data from the above papers, we
determined the evolution parameters of the orbital
period and refined the rate of its change. In addition,
such a large set of observational data allowed us to
study in detail the precessional variability.


\begin{table*}[t]
\centering
{{\bf Table 1} Times of superorbital maxima for LMC X-4 from the BAT/SWIFT data}\label{eclipses}
\vspace{5mm}\begin{tabular}{c|c|c|c} \hline\hline
{Time of maximum} & {Time of maximum} & {Time of maximum} & {Time of maximum}\\
{(MJD)}    & {(MJD)}    & {(MJD)}    & {(MJD)}  \\
\hline
$53442.083(262)$ & $54379.497(221)$ & $55322.751(278)$ & $56264.192(212)$ \\
$53472.105(132)$ & $54413.221(041)$ & $55351.775(278)$ & $56295.253(149)$ \\
$53502.218(279)$ & $54444.073(163)$ & $55382.556(267)$ & $56325.917(172)$ \\
$53536.030(279)$ & $54474.505(197)$ & $55412.570(267)$ & $56356.301(251)$ \\
$53562.919(279)$ & $54505.674(200)$ & $55444.178(267)$ & $56387.028(201)$ \\
$53593.763(175)$ & $54535.892(181)$ & $55474.534(190)$ & $56417.406(201)$ \\
$53624.001(134)$ & $54565.828(251)$ & $55505.001(159)$ & $56448.498(239)$ \\
$53653.959(207)$ & $54596.526(134)$ & $55534.980(239)$ & $56478.777(333)$ \\
$53684.864(333)$ & $54625.600(263)$ & $55566.299(239)$ & $56509.446(247)$ \\
$53714.052(223)$ & $54656.823(198)$ & $55596.410(270)$ & $56539.627(247)$ \\
$53744.880(223)$ & $54687.489(333)$ & $55626.959(199)$ & $56570.707(223)$ \\
$53776.377(223)$ & $54718.591(152)$ & $55659.240(199)$ & $56600.626(167)$ \\
$53806.366(223)$ & $54749.088(168)$ & $55688.834(239)$ & $56631.122(199)$ \\
$53835.519(168)$ & $54778.980(131)$ & $55718.794(239)$ & $56662.332(243)$ \\
$53866.359(180)$ & $54809.656(261)$ & $55750.048(239)$ & $56691.965(243)$ \\
$53896.825(253)$ & $54839.357(221)$ & $55780.720(233)$ & $56722.428(197)$ \\
$53927.010(269)$ & $54869.611(137)$ & $55810.834(262)$ & $56753.351(256)$ \\
$53957.470(118)$ & $54899.375(189)$ & $55841.061(248)$ & $56783.600(198)$ \\
$53988.321(097)$ & $54930.155(175)$ & $55872.187(204)$ & $56814.124(198)$ \\
$54019.099(098)$ & $54960.006(172)$ & $55902.129(183)$ & $56844.112(333)$ \\
$54049.203(137)$ & $54989.646(207)$ & $55932.555(207)$ & $56873.079(214)$ \\
$54079.558(147)$ & $55020.347(333)$ & $55962.561(186)$ & $56905.167(214)$ \\
$54109.563(192)$ & $55051.161(232)$ & $55993.282(244)$ & $56934.614(245)$ \\
$54140.746(248)$ & $55080.931(277)$ & $56023.652(255)$ & $56965.146(203)$ \\
$54170.838(180)$ & $55110.628(128)$ & $56053.418(247)$ & $56994.703(244)$ \\
$54202.570(238)$ & $55141.687(136)$ & $56082.717(247)$ & $57025.923(249)$ \\
$54231.799(219)$ & $55171.818(223)$ & $56113.865(229)$ & $57056.006(258)$ \\
$54262.646(219)$ & $55201.876(166)$ & $56143.784(232)$ & $57086.687(278)$ \\
$54291.012(219)$ & $55232.672(259)$ & $56173.306(221)$ & $57116.417(218)$ \\
$54319.538(181)$ & $55261.729(208)$ & $56203.936(178)$ & $57146.931(218)$ \\
$54352.065(135)$ & $55292.334(208)$ & $56233.660(233)$ & $$ \\
\hline
\end{tabular}
\end{table*}


\section*{Observations and data analysis}

To investigate the 30-day modulation of the flux
from LMC X-4, we used all the available data
from wide-field X-ray space telescopes performing a
quasi-continuous monitoring (the source is almost
always observed every day) of the entire celestial
sphere.

To find the precessional variability parameters, we
used the data from the BAT (Burst Alert Telescope,
Krimm et al. 2013) telescope of the SWIFT observatory
(Gehrels et al. 2004) that were obtained in
the 15-50 keV energy band and are publicly available
(http:// swift.gsfc.nasa.gov/results/transients/LMCX-4/).
The light curves used have a time resolution of
$\sim90$ min and span the time interval from February 2005
to May 2015. The telescope is sensitive enough
to be able to determine the position of each
maximum of the 30-day flux modulation ``wave'' on
the time scale of all 10 years. For this purpose, we
fitted the light curve near the maxima by Gaussians
$C(t)=N~e^{-\frac{(t-T_{\psi_0}^i)^2}{2\sigma^2}}$,
restricted by the time interval
[$T_{\psi_0}^i-10$,$T_{\psi_0}^i+10$] days, where $T_{\psi_0}^i$
is the position of the i-th maximum. Thus, we found the time vector
$\overrightarrow{T_{\Psi_0}}=[T_{\Psi_0}^1, T_{\Psi_0}^2,...,T_{\Psi_0}^{123}]$
of the maxima of the 30-day cycle with the corresponding confidence intervals
($\simeq0.2$ day is a typical value, see Table 1). Figure 1
shows that the light curve is adequately fitted by these
Gaussians near the maxima in an arbitrarily taken
time interval.

To check whether the solution for the precessional
modulation obtained in the 15-50 keV X-ray band
was applicable to the measurements in the soft X-ray
band (2-20 keV), we used data from the Japanese
MAXI all-sky monitor (Matsuoka et al. 2009) installed
on the Japanese experimental module of the
International Space Station. The monitor has been
observing LMC X-4 since August 2009 and, thus,
allows an independent time series of data completely
overlapping in time with the BAT/Swift data to be
obtained. The MAXI data are publicly available at
http://maxi.riken.jp/top/.

We also retrospectively applied our 30-day modulation
model to the monitoring data for LMC X-4
in the 20-70 keV X-ray energy band obtained from
June 1991 to June 2000 with the BATSE/Compton-GRO telescope
(Gehrels et al. 1993) and to the data
in the 2-12 keV energy band obtained from January
1996 to December 2011 with the All-SkyMonitor
(ASM) (Levine et al. 1996) onboard the
RXTE observatory (Bradt et al. 1993).

To find the X-ray eclipses associated with the orbital
motion, we used all of the available open data
from the IBIS telescope (Ubertini et al. 2003) of
the INTEGRAL gamma-ray observatory (Winkler
et al. 2003), the Proportional Counter Array (PCA)
(Jahoda et al. 2006) of the RXTE observatory, and
the EPIC camera (Struder et al. 2001) of the XMM-Newton
orbital observatory.

We analyzed the IBIS/INTEGRAL data based
on the balanced cross-correlation algorithm (for a
description, see Krivonos et al. 2010; Churazov
et al. 2014).

\begin{figure}
\begin{center}
\includegraphics[width=0.95\columnwidth,bb=60 180 560 675,clip]{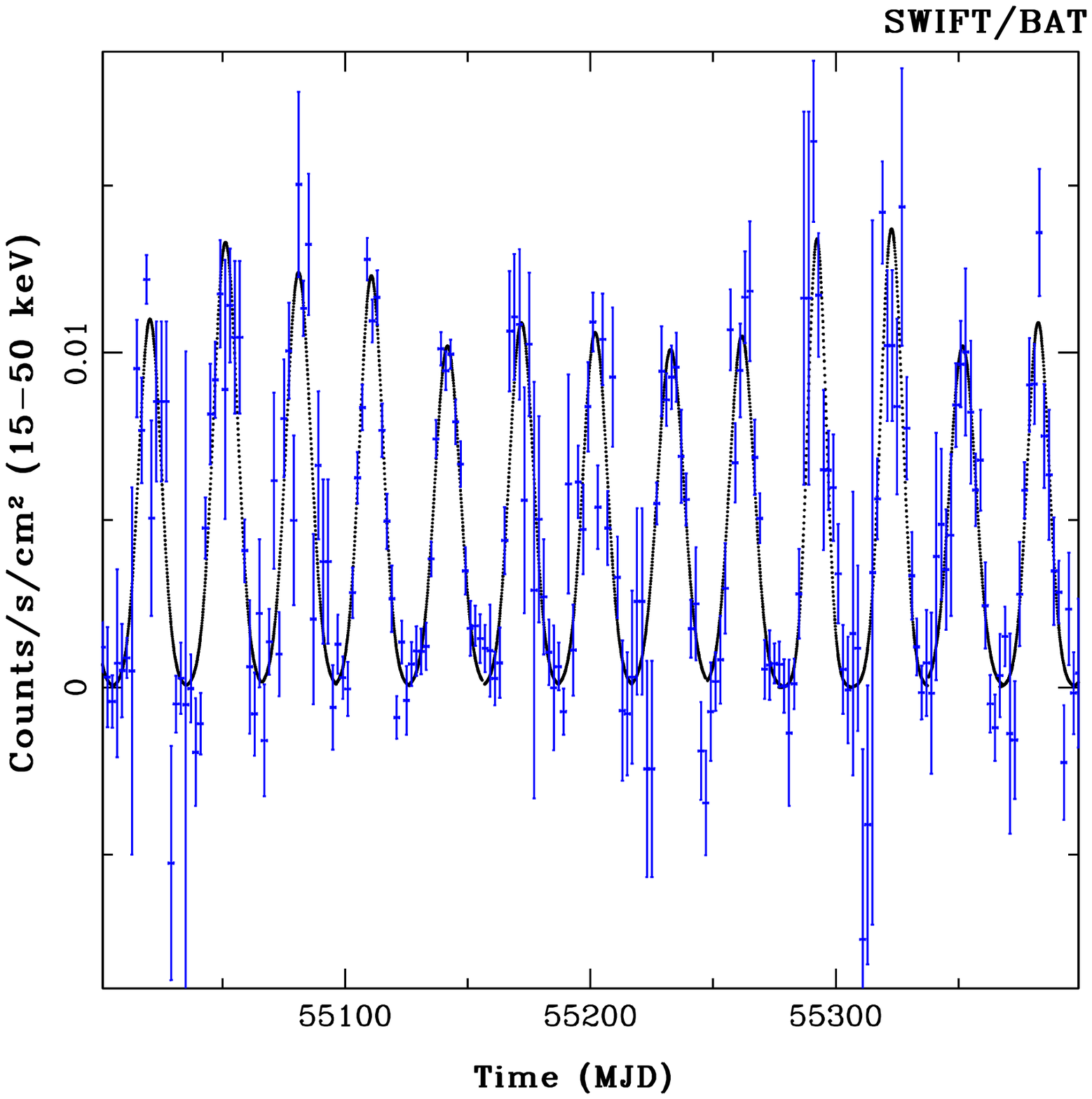}
\end{center}
\caption{An example of fitting the light curve of the X-ray pulsar LMC X-4 by
a sequence of Gaussians near the maxima of the 30-day cycle (solid line).
The light curve was constructed in the 15-50 keV energy band from the BAT/SWIFT data.}
\label{fig1}
\end{figure}

\section*{Results}

As has been pointed out above, in this paper we
consider questions related to the long-term variability
of the X-ray flux from LMC X-4. Therefore, the
results obtained were divided into two parts, those
associated with the precessional and orbital motions.

\subsection*{Precessional Variability}

The long-term variability in some X-ray binaries
has also been investigated previously (see, e.g.,
Clarkson et al. 2003; Kotze and Charles 2012; and
references therein). It was established that among all
of the sources exhibiting such variability, LMC X-4
has the most stable precession period. However,
all of these results were obtained under conditions
of limited statistics: despite the fact that the
source's luminosity is very high
($\sim 10^{38}$ erg s$^{-1}$), it
is at a distance of 50 kpc away from us, and its
flux near the Earth does not exceed a few
mCrab in soft X-rays. Therefore, it was not possible
to determine the time of each maximum of the 30-day
cycle, and the epoch-folding technique, where
several cycles are averaged to improve the statistics,
which reduces considerably the number of degrees of
freedom (independent measurements), was applied to
determine the precessional ephemerides. In addition,
the data in the soft 2-12 keV X-ray band were used in
the above papers, while the maximum of the emission
detected from LMC X-4 occurs at an energy $>20$ keV
(Tsygankov and Lutovinov 2005).

As has been pointed out above, the BAT/Swift
telescope is sensitive enough to trace the evolution of
the flux from the source, and we found the time vector
of the maxima of the 30-day cycle. Next, to analytically
approximate the precession cycle, we assumed
it to have a periodic pattern. Denoting the phase at
which the maximum on the light curve is reached by
$\Psi_0=0$ (phase ``0''), we can then determine the phase
at some arbitrary time t from the formula

$$ \Psi(t)=\left\{\frac{(t-T_0)}{P_{sup}}-\frac{(t-T_0)^2~\dot{P_{sup}}}{2~P_{sup}^2}\right\}\hspace{1cm}(1)$$

where the time  $T_0$ corresponds to the phase $\Psi_0=0$, $P_{sup}$
is the precession period at the time $T_0$; $\dot{P_{sup}}$ is
the first derivative of the period at the time $T_0$ (in
the above expression we will restrict ourselves only to it),
and the braces denote the operation of taking the
fractional part (the integer part is the ``epoch'' at time
``t''). These three parameters are the ephemerides of
the precessional modulation. We can find their best
values by applying the $\chi^2$ test and minimizing the
functional
$\sum_{i=1}^{i=123}\frac{\Psi(T_{\Psi_0}^i)^2}{\delta\Psi(T_{\Psi_0}^i)^2}$.
We obtained the best $\chi^2$ value for the following set of parameters:
$T_0=53441.50\pm0.03$ MJD, $P_{sup}=30.370\pm0.001$ days,
$\dot{P_{sup}}=(0\pm1)\times10^{-5}$. Nevertheless, the $\chi^2$ value
is considerably larger than unity, which may suggest
that either we underestimated the errors in the positions
of the maxima of the 30-day modulation wave
or our hypothesis (1) is not quite correct.

\begin{figure}
\begin{center}
\includegraphics[width=0.95\columnwidth,bb=60 180 560 675,clip]{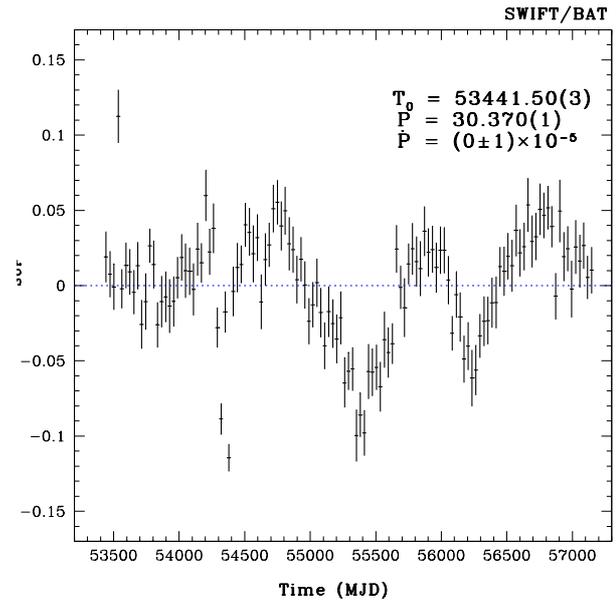}
\end{center}
\caption{Phase shift of the precessional modulation from the best solution obtained
over the entire period of BAT/Swift observations (for more details, see the text).}
\label{fig2}
\end{figure}

Figure 2 shows the deviations of the precession
phase for the vector $\overrightarrow{T_{\Psi_0}}$ 
from the best solution obtained
for the entire set of BAT/SWIFT observational
data. It follows from the figure that this solution
describes the precessional periodicity, on average,
satisfactorily; at the same time, deviations, both in
individual measurements and systematic ones, that
exceed the admissible values for the normal distribution
are observed. This may suggest that the
variability in the binary is more complex than the
model (1) we proposed, and there exists, for example,
a time dependence of the ephemerides. Figure 2
shows that this dependence must be fairly complex
and nonlinear in pattern; in particular, including the
second derivative in the model does not improve the
situation. Nevertheless, it should be noted that the
parameters we obtained can be used to describe the
observational data.

\begin{table*}[t]
\centering
{{\bf Table 2.} Parameters to calculate the superorbital phase correction for the ephemerides
$T_0=53441.53383$~MJD and $P_{sup}=30.370$ days}\label{tbl2}

\vspace{5mm}\begin{tabular}{c|c|c|r|r} \hline\hline
  $~~~~~~$    & $~~~~~~~~~~~~~$ & $~~~~~~~~~~~~~$ & $~~~~~~~$ & $~~~~~~~~~~$\\
{j}& $T_1$ & $T_2$ & $\Psi_c~~~$ & $\dot{\Psi_c}\times 10^{-4}$\\
  $$    & (MJD) & (MJD) & $$ & $$\\
\hline
$1$   &  $45000.0$   & $51100.0$  &  $ 0.190$ & $ 0.000$ \\
$2$   &  $51100.0$   & $52400.0$  &  $ 0.190$ & $-1.692$ \\
$3$   &  $52400.0$   & $53900.0$  &  $-0.030$ & $ 0.000$ \\
$4$   &  $53900.0$   & $54200.0$  &  $-0.030$ & $ 1.666$ \\
$5$   &  $54200.0$   & $54400.0$  &  $ 0.020$ & $-3.000$ \\
$6$   &  $54400.0$   & $54700.0$  &  $-0.040$ & $ 2.333$ \\
$7$   &  $54700.0$   & $55400.0$  &  $ 0.030$ & $-2.000$ \\
$8$   &  $55400.0$   & $55950.0$  &  $-0.110$ & $ 2.545$ \\
$9$   &  $55950.0$   & $56250.0$  &  $ 0.030$ & $-3.166$ \\
$10$  &  $56250.0$   & $56750.0$  &  $-0.065$ & $2.200$ \\
$11$  &  $56750.0$   & $57000.0$  &  $ 0.045$ & $-1.800$ \\
$12$  &  $57000.0$   & $58000.0$  &  $ 0.000$ & $0.000$ \\
\hline
\multicolumn{5}{l}{}\\ [-3mm]
\end{tabular}
\end{table*}

To check how applicable our solution is for a wider
time interval and for other energy bands, we analyzed
the light curves of LMC X-4 obtained with other
X-ray instruments having long-term observations of
this binary. In particular, we used the ASM/RXTE
and MAXI data for the soft X-ray energy band (2-12 and 2-20 keV,
respectively) and the BATSE data
for the 20-70 keV energy band. The data from all
three instruments are statistically not good enough
to determine the time of each maximum of the 30-day
modulation from them, as was done using the
BAT/SWIFT data. Therefore, to improve the statistical
significance of the signal, we used the epochfolding
technique and averaged several periods of the
precession cycle, thereby passing from the temporal
light curves to the light curves on the phase plane, and
determined the phase shift in each of the averagings
(foldings). For different instruments, we took different
``windows'' for folding and different successive shifts
of the window: 400 and 60 days for BATSE, 200 and
60 days for ASM, 100 and 30 days for MAXI, respectively.
We used a ``shift'' smaller than the window
duration to trace in more detail the possible changes,
although it should be noted that the measurements
cease to be independent in such an approach. As
a result, we obtained the phase shifts of the 30-day
cycle relative to the presumed model as a function of
time for a 25-year interval of observations (the left
panel in Fig. 3). It follows from the figure that the
period in the early observations was also approximately
constant and close to the value measured from the BAT data,
but its phase was shifted significantly.
The subsequent evolution of the phase shift has no
stable trend but rather changes chaotically. Possible
explanations of such a behavior are discussed in the
concluding part of the paper.

To compare the phase changes in the soft and
hard X-ray bands, we used the MAXI data in the
2-20 keV energy band and the simultaneous BAT
observations in a harder X-ray band (15-50 keV). It
is clearly seen from Fig.3 that the phase changes in
both energy bands occur synchronously, from which
it follows that the data obtained in different X-ray
bands for the source can be simultaneously used to
investigate the long-term variability. In addition,
this fact suggests that the observed precession phase
variations are real and cannot be explained by the
measurement errors.

To investigate the dependence of the emission
characteristics for LMC X-4 on the phase of the 30-day
cycle, it is important to be able to determine it
as accurately as possible. Obviously, the dependence
presented in Fig.3 cannot be described by a simple
analytical model for the entire interval of observations
under consideration. Therefore, we propose to use
the model with the constant precession
period determined above and the tabulated phase
corrections. Thus, the phase of the 30-day cycle can
be calculated in general form from the formula

$$ \Psi(t)=\left\{\frac{t-T_0}{P_{sup}}\right\}-(\Psi_c^j+\dot{\Psi_c^j}*(t-T_1^j)), ~t \in [T_1^j,T_2^j] \hspace{0.1cm}(2)$$

where $P_{sup}=30.370$ days and $T_0=53441.50$ MJD,
the braces denote the operation of taking the fractional part, and
the correction coefficients and the corresponding time
intervals are given in Table 2. The right part of Eq. (2)
(phase correction) is indicated in Fig.3 by the solid
line. The right panel in the same figure presents the
deviation of the precessional modulation phase ``0''
obtained after the data correction in accordance with
Eq. (2). It can be seen that after such a correction,
the residual variations of the zero phase deviation are
within approximately 5\%.

\begin{figure}
\begin{center}
\includegraphics[width=0.99\columnwidth,bb=0 200 575 575,clip]{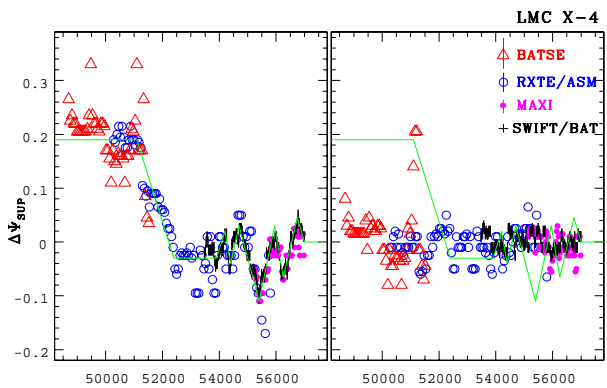}
\end{center}
\caption{Deviations of the phase of the maxima of the 30-day cycle from the expected value
versus time for the ephemerides $T_{sup,0}=53441.50$~MJD and $P_{sup}=30.370$~days.
The data were obtained over almost 25 years of observations with four
instruments. The solid line indicates an empiricalmodel describing the phase
change (see also Table 2). The left panel presents the directmeasurement results;
the right panel presents those corrected for thismodel (see the text).
Typical errors are indicated by vertical bars on the symbols in front of
the instrument name.}
\label{fig3}
\end{figure}

Figure 4 shows the average flux profiles for LMC X-4
as a function of the precession phase calculated
from Eq. (2) for four instruments operating in different
energy bands. The presented profiles are similar
to one another in shape and are symmetric relative
to their maxima (for the convenience of perception,
phase ``1'' corresponds to the maximum). It can
also be noted that when using mCrab as a unit
of flux, the source in hard X-rays turns out to be
significantly brighter than in soft X rays. The latter
stems from the fact that the spectrum of LMC X-4 differs
significantly from the spectrum of the Crab Nebula.

\subsection*{The Model of Orbital Motion}

The orbital period of an X-ray binary system harboring a
pulsar can be determined most accurately by measuring
the change in the spin period of the neutron
star with orbital phase (based on the Doppler effect).
The period obtained in this way was called the meanlongitude
period $P_{\pi/2}$ in the literature, and the meanlongitude
time $T_{\pi/2}$ (see, e.g., Smart 1953) is taken as
the initial time (``0'') in this case. For high-inclination
binaries, in which eclipses of the X-ray emission originating
near the compact object by the normal star
are observed, the orbital period can be determined
from the frequency of such eclipses. The period and
initial time obtained in this way are called ``eclipsing''
ones and are denoted by $P_{ecl}$ and $T_{ecl}$, respectively.
Generally, both period/time pairs are related between
themselves in a complex nonlinear way (see, e.g.,
Smart 1953). A sufficient condition for the equalities
$P_{\pi/2}=P_{ecl}$ and $T_{\pi/2}=T_{ecl}$ is a circular orbit in the
binary (in other words, with the eccentricity $e=0$).

\begin{figure}
\begin{center}
\includegraphics[width=0.99\columnwidth,bb=0 200 575 675,clip]{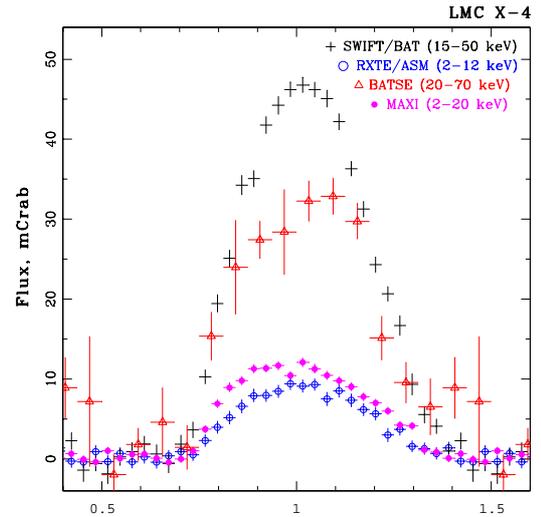}
\end{center}
\caption{Dependence of the flux recorded from the X-ray pulsar LMC X-4 in different energy bands
on precession (superorbital) phase from the data of four X-ray instruments. The folding
was performed with the period $P_{sup}=30.370$~days and by applying
the correction for the historical ``walk'' of the phase of maximum (Eq. (2)).}
\label{fig4}
\end{figure}

The measurements made precisely by monitoring
the changes in the phase and/or period of the neutron
star spin were used in most previous papers aimed
at determining the parameters of the orbital motion
in the binary (for brevity, we will call this method
``based on the Doppler effect''; see references in Table
3). However, such an approach is rather costly
from the standpoint of organizing observations and
the requirements imposed on them. First, the intensity
of LMC X-4 is low, and sensitive instruments
capable of recording its emission at a statistically
significant level in relatively short time intervals are
needed. Second, a good time resolution is needed to
measure the pulsation period and its changes. Third,
the series of observations must be compact in time
with a total exposure time near or slightly longer
than the orbital period, $\sim1.4$ days. Therefore, there
were few such measurements over the entire history
of observations of LMCX-4, and they were performed
with a low cadence in time. Nevertheless, they turned
out to be sufficient to detect a decrease in the orbital
period in the binary (see, e.g., Levine et al. 2000, and
references therein).

\begin{figure}
\begin{center}
\includegraphics[width=0.99\columnwidth,bb=0 200 575 675,clip]{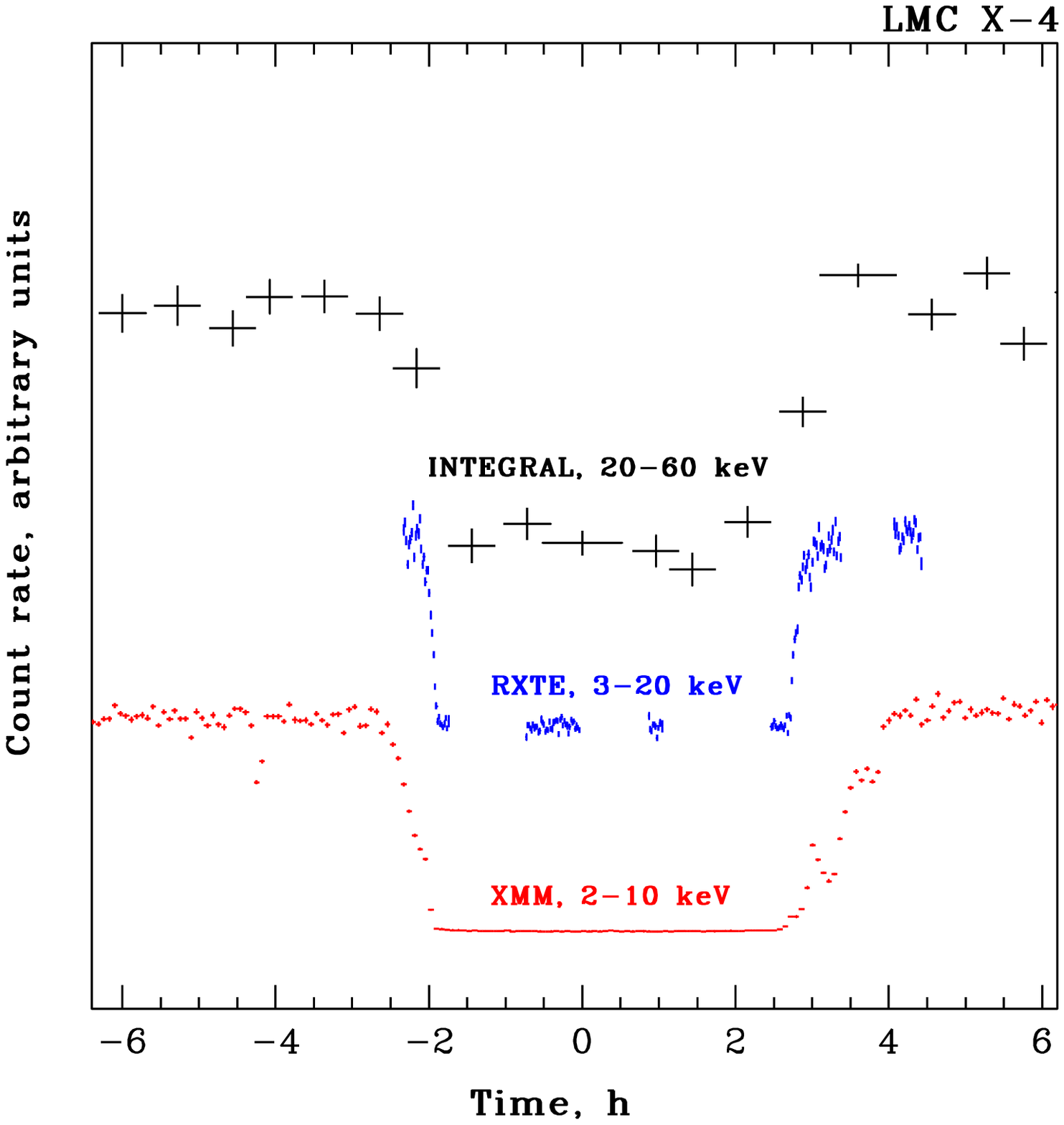}
\end{center}
\caption{
Examples of the profile of an orbital X-ray eclipse in LMC X-4 from the data of three
instruments in different energy bands.}
\label{fig5}
\end{figure}

The model of orbital motion can also be constructed
by using the data only on the eclipse times.
In this case, it is natural that the more eclipses were
registered and the wider the interval of observations,
the more accurately we can determine the orbital parameters
and trace their evolution (see, e.g., Falanga
et al. 2015). The last decades have been marked
by the appearance and successful in-orbit operation
of many sensitive X-ray instruments, which have
accumulated a large set of observational data for
various celestial objects, including those for LMC X-4,
with its total exposure time exceeding several
million seconds. Taking into account the fairly short
orbital period in the binary, dozens of orbital eclipses
must be recorded in such a time. We searched
for them in the data from three instruments over
the last 19 years starting from 1996: PCA/RXTE,
IBIS/INTEGRAL, and XMM-Newton. Since these
instruments operate in different energy bands, we
primarily made sure that the eclipses recorded by
them had a similar shape. Figure 5 shows the profiles
of three eclipses registered by the above instruments.
It can be seen that the duration of a total eclipse does
not depend, with a good accuracy, on the energy band,
and only the steepness of the eclipse ingress and
egress slightly differs from instrument to instrument.
In addition, we made sure that the eclipse shape is
also retained for the eclipses registered at different
times with the same instrument. Thus, we found
43 eclipses in the available observational data.
The mid-eclipse times ($T_{ecl}$) and the corresponding measurement
errors (at the $1\sigma$ level)are given in Table 3.
Since the orbital eccentricity in LMC X-4 is close to
zero (and, consequently, $T_{ecl}=T_{\pi/2}$), we added the
measurements of the times $T_{\pi/2}$ found in the literature
(nine more measurements, see Table 3) to our times $T_{ecl}$,
and constructed a model of the orbital period
change over almost 40 years of observations from the
entire series.

\begin{figure}
\begin{center}
\includegraphics[width=0.9\columnwidth,angle=-90,bb=90 80 730 950,clip]{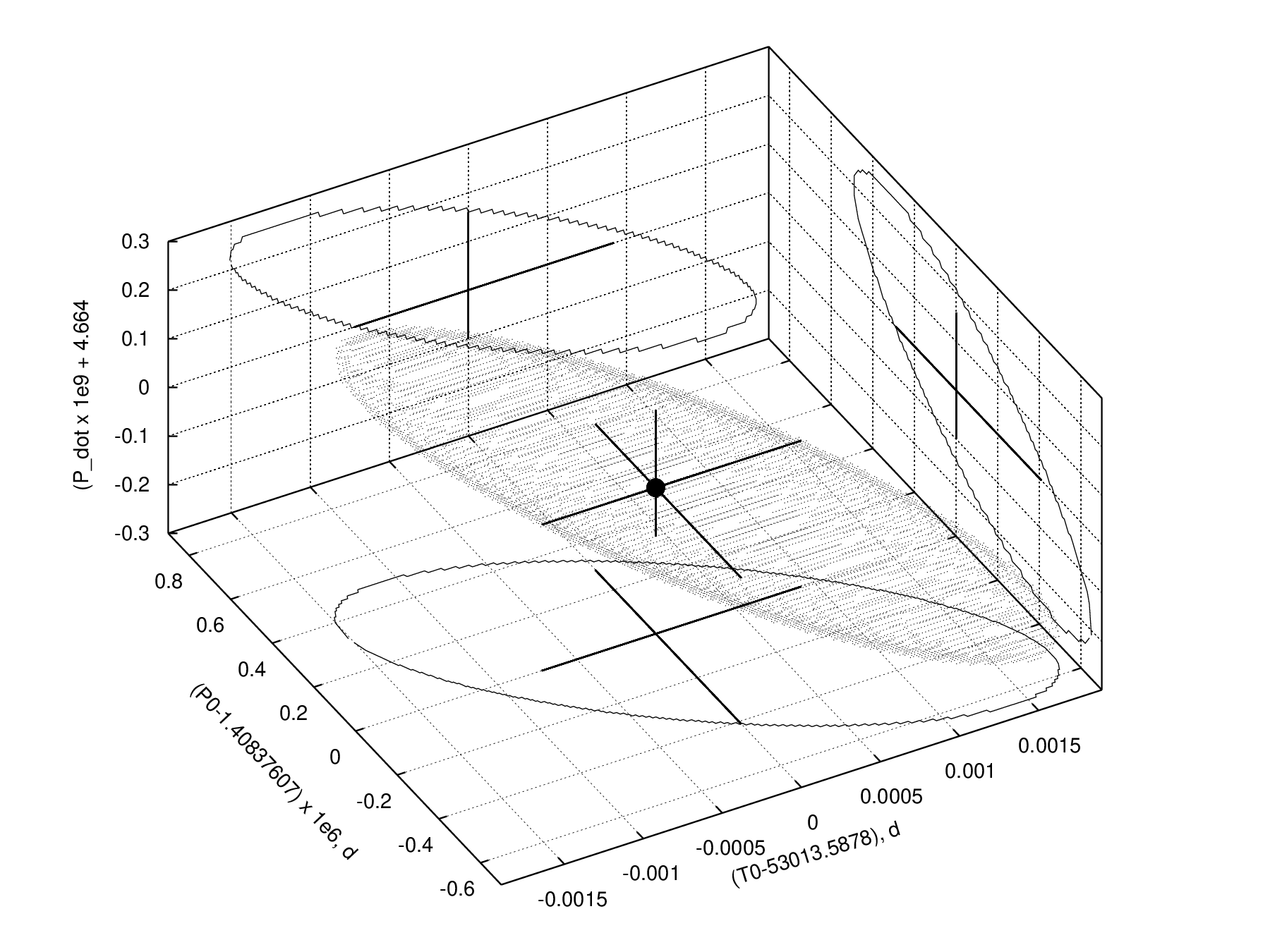}
\end{center}
\vspace{-1cm}
\caption{($1\sigma$) error region for the best-fit parameters describing the orbital
motion in LMC X-4.}
\label{fig6}
\end{figure}

To determine themodel parameters, we applied the
same method as previously when finding the solution
for the 30-day cycle, i.e., we used Eq.(1) to find the
phase and  $\overrightarrow{T_{ecl,\pi/2}}$ as the measurement
vector. In this case, the model describes satisfactorily the
observational data ($\chi^2\simeq80$ for 49 degrees of freedom), and
we obtained the following parameters of the model
of orbital motion:
$T_0^{orb}=53013.5878^{+0.0009}_{-0.0007}$~MJD,
$P_0^{orb}=1.40837607^{+2.9\cdot 10^{-7}}_{-4.1\cdot 10^{-7}}$ days
and $\dot{P}_0^{orb}=(-4.66^{+0.16}_{-0.10})\times10^{-9}$.
The presented errors correspond to the $1\sigma$ deviation under
the assumption that all parameters are independent. The latter is not
obvious; for example, an ``underestimated'' period can
be ``compensated for'' by an increase in the rate of
its change, i.e., these parameters can be correlated.
To estimate the degree of correlation between the
parameters and to obtain more reliable errors in the
parameters, we calculated the $1\sigma$ deviations in the
three-dimensional space ($T_0^{orb}, P_0^{orb}, \dot{P}_0^{orb}$)
of errors. The region of admissible values and its projections
onto the corresponding planes of parameters are
shown in Fig. 6. For comparison, the line segments
indicate the errors calculated by assuming the parameters 
to be independent. It follows from the figure that
the parameters have a significant correlation between
themselves and that the above errors were greatly
underestimated. Given all of the aforesaid, the final
solution for orbital motion appears as follows:\\
\hspace{2cm}$\mathbf{T_0^{orb}=53013.5878^{+0.0018}_{-0.0015}}$~MJD\\
\hspace{2cm}$\mathbf{P_0^{orb}=1.40837607^{+4.9\cdot 10^{-7}}_{-6.5\cdot 10^{-7}}}$ days\\
\hspace{2cm}$\mathbf{\dot{P}_0^{orb}=(-4.66\pm0.26)\times 10^{-9}}$.\\
The latter value can be transformed to a more customary
and convenient form for comparison with the results
of previous measurements,
$\dot{P}_0^{orb}/P_0^{orb}=(-1.21\pm 0.07 )\times 10^{-6}$ yr$^{-1}$.
Comparison with the results of previous measurements
$\dot P_{orb}/P_{orb}=(0.98\pm0.07)\times10^{-6}$ yr$^{-1}$ 
(Levine et al. 2000) shows that the decay rate in the orbital period
in LMC X-4 is actually higher than has been thought previously.

\begin{figure}
\begin{center}
\includegraphics[width=0.99\columnwidth,bb=0 200 575 675,clip]{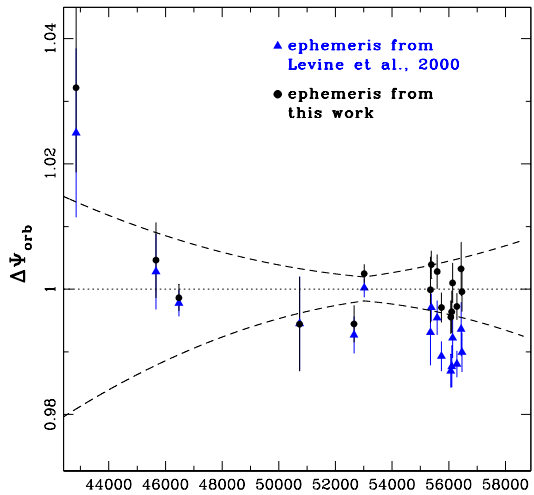}
\end{center}
\caption{
Comparison of the quality of the applicability of two groups of orbital ephemerides in
LMC X-4 to the observational data: the triangles -- Levine et al. (2000);
the circles -- this paper. The dashed lines indicate the $3\sigma$ error region for
the solution obtained in this paper.}
\label{fig7}
\end{figure}

Figure 7 shows the dependence of the orbital
phase deviation from zero value (by definition, phase
``0'' correspond to the mid-eclipse) calculated from
Eq.(1) with the orbital motion parameters obtained
above (filled circles). To improve the visual perception
of the figure, the data were grouped in time with a
30-day step. The dashed lines bound the $3\sigma$ error
region within which all of the measurement results
lie. For comparison, the triangles in the same figure
indicate the orbital phase deviations obtained by using
the orbital ephemerides of LMC X-4 from Levine
et al. (2000). It can be seen that our model describes
the orbital motion in LMC X-4 considerably better,
but it was constructed fromthe data spanning amuch
longer time interval. Such a significant increase
of the observational data set is associated mainly
with the series of deep observations of the Large
Magellanic Cloud performed by the INTEGRAL
observatory in 2003-2013 (Grebenev et al. 2013).
Interestingly, the decay rate in the orbital period
$\dot P_{orb}/P_{orb}=(1.00\pm0.05)\times10^{-6}$ yr$^{-1}$
obtained by Falanga et al. (2015), who used only the data from the
first series of INTEGRAL observations of LMC X-4
in 2003-2004 in addition to the historical data,
turns out to be similar to the results from Levine et al.
(2000). The latter may suggest a change in the decay rate
in the orbital period in the binary in the last
decade, although the significance of these differences
is low (about $2\sigma$).

Applying the solution obtained for the orbital motion
in LMC X-4, we constructed an averaged
profile of the dependence of the flux registered
from the source on orbital phase from the data of
four instruments operating in different X-ray bands
(Fig. 8). To increase the statistical significance of
the signal, we used only the data obtained near the
maxima of the 30-day cycle $\Psi_{sup} \in [0.8-1.2]$. The
averaged profiles constructed in this way in different
energy bands are similar to one another in shape, as
we have pointed out for the profiles of single eclipses.

\section*{Discussion}

We presented the results of a comprehensive study
of the long-term variability of the X-ray flux from
the high-mass X-ray binary LMC X-4 based on the
data of several space observatories over the last 25 years.

\begin{figure}
\begin{center}
\includegraphics[width=0.99\columnwidth,bb=0 200 575 675,clip]{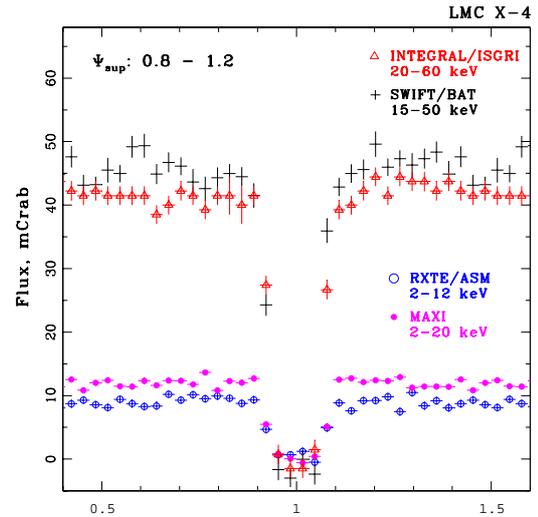}
\end{center}
\caption{
Averaged dependence of the X-ray flux recorded from LMC X-4 on orbital phase from the data
of four instruments. To increase the statistical significance, we used only the data for
the range of precession phases $\Psi_{sup}=0.8-1.2$.}
\label{fig8}
\end{figure}

We obtained the mean precession period $P_{sup}=30.370$ days,
which may be considered as the equilibrium
one over the time interval under consideration.
Note significant variations of the period measured
in each specific cycle near this value, causing the
measured phase of the 30-day period to ``walk'' with
respect to the phase predicted by the model with a
constant period. The shape of the ``walk'' is very
similar to the shape of the deviation of some quantity
from its initial value in a ``Brownian'' process.
Indeed, if we represent the expression to calculate
the times of some phase in successive cycles as
$T_i=T_0+\sum_{j=0}^{j=i}P_j$, where i is the cycle number and
$P_j$ is the period of the jth cycle, and if assume that
[$P_j$] obey a normal distribution, then [$T_j$] will have the
frequency characteristic of ``red noise'' (corresponding
to Brownian motion), which, by definition, is a
``white noise'' integral. Unfortunately, because of the
insufficient quantity and quality of observational data,
we cannot test this hypothesis by constructing the
power spectrumof the time series. Nevertheless, if the
suggested hypothesis is valid, then it is necessary to
determine the processes responsible for the changes
in the precession period. Two processes that explain
the existence of the 30-day modulation are usually
considered: tilted accretion disk precession or
radiation-induced accretion disk warping. Neither of
the two scenarios forbids the period variability, although
they do not give a clear answer to the question
of why and how this can occur.

Using long-term observations, we refined the orbital
period in the binary system and its decay rate. In
particular, this decay rate in the orbital period
$\dot{P}_0^{orb}/P_0^{orb}=(-1.21\pm 0.07 )\times 10^{-6}$ yr$^{-1}$
turned out to be 20\% higher than the values obtained
previously (Levine et al. 2000; Falanga et al. 2015).
The latter can be connected with a change in the decay rate
in the orbital period in the last decade.

In general, investigating the orbital period changes
can give valuable information about the evolution
of close binaries and the mechanisms responsible
for this process. For LMC X-4, the current orbital
period has been measured many times: from the
SAS-3 (Kelley et al. 1983), EXOSAT (Dennerl 1991),
Ginga, RXTE (Levine et al. 1991, 2000), ROSAT
(Levine 1996), and INTEGRAL (the first series of
observations in 2003-2004; Falanga et al. 2015)
data. Levine et al. (1991) were the first to point to
a decrease in the orbital period in the binary and
to obtain an upper limit on its derivative. Subsequently,
using additional data, Safi-Harb et al. (1996)
and Levine (2000) confirmed these conclusions.
LMC X-4 is one of the five high-mass X-ray binaries
(HMXBs) for which significant decreases in
the orbital periods have been detected: LMC X-4,
Cen X-3, 4U 1700-377, SMC X-1, and OAO 1657-415
(see the review of Falanga et al. 2015 and
references therein). Various models were proposed
to explain the observed changes in the orbital period
(Kelley et al. 1983; van der Klis 1983, 1984; Levine
et al. 1993, 2000; Rubin et al. 1996; Safi-Harb et al.
1996; Jenke et al. 2011). They are all based to
some extent on two mechanisms that can lead to
a decrease in the orbital period in the binary: tidal
interactions and mass transfer due to a fast stellar
wind, which characterizes most of the companion
stars in HMXBs, or evolution of the companion star
itself, in particular, its expansion and an increase
in the moment of inertia, causing the star to spin
down. The tidal angular momenta will then transfer
this change to the orbital angular momentum to
synchronize the binary, causing the orbital period to
decrease (see, e.g., Levine et al. 1993, and references
therein).

Our calculations show that both tidal interactions and mass
transfer can play an important role for such a close
binary as LMC X-4. The latter, in turn, must lead to
a change in the accretion rate onto the neutron star
and, as a consequence, to a change in its luminosity.
This must be particularly noticeable in the last
decade, when an increase in the rate of decrease in
the orbital period is observed. Based on INTEGRAL
data, we attempted to detect possible trends in the
X-ray flux from LMC X-4, while the X-ray flux in
such binaries must correlate with the mass transfer
rate. For this purpose, we used only out-of-eclipse
data near the precessional modulation maxima. From
these results we concluded that, within the measurement
error limits, the flux may be considered constant.
Taking this into account, the model predicting
a change in the orbital period due to the evolution of
the normal star may be more plausible for the binary
under consideration (recall that an O8 III giant is the
companion star in it). To confirm or refute this hypothesis,
it is necessary to perform optical monitoring
observations and to compare them with the archival
data, while its confirmation can become the first direct
argument for the model of a change in the orbital
period in the binary due to an evolving and expanding
normal star.

\section*{ACKNOWLEDGMENTS}

{\sl 
This work was financially supported by the Russian
Foundation for Basic Research (project no. 13-02-12094
and grant President of Russian Federation NSh 6137.2014.2).
We are grateful to E.M. Churazov, who
developed the IBIS/INTEGRAL data analysis methods
and provided the software. We thank the International
Space Science Institute (Bern, Switzerland)
for organizing a meeting on the theme of the publication.
}

\bigskip

\section*{References}
\parindent=0mm

1. H. V. Bradt, R. E. Rothschild, and J. H. Swank,
Astrophys. J. Suppl. Ser. 97, 355 (1993).

2. C. Chevalier and S. A. Ilovaisky, Astron. Astrophys.
59, L9 (1977).

3. E. Churazov, R. Sunyaev, J. Isern, J. Knodlseder,
P. Jean, F. Lebrun, N. Chugai, S. Grebenev, et al.,
Nature 512, 406 (2014).

4. W. I. Clarkson, P. A. Charles,M. J. Coe, and S. Laycock,
Mon. Not. R. Astron. Soc. 343, 1213 (2003).

5. K. Dennerl, PhD Thesis (Max Planck Inst. Extraterrestr.
Phys., Univ.Munchen, 1991).

6. A. Epstein, J. Delvaille, H. Helmken, S. Murray,
H. W. Schnopper, R. Doxsey, and F. Primini, Astrophys.
J. 216, 103 (1977).

7. M. Falanga, E. Bozzo, A. Lutovinov, J. M. Bonnet-
Bidaud, Y. Fetisova, and J. Puls, Astron. Astrophys.
577, 16 (2015).

8. N. Gehrels, E. Chipman, and D. A. Kniffen, Astrophys.
J. Suppl. Ser. 97, 5 (1993).

9. N. Gehrels, G. Chincarini, P. Giommi, K. D. Mason,
J. A. Nousek, A. A. Wells, N. E. White,
S. D. Barthelmy, et al., Astrophys. J. 611, 1005
(2004).

10. R. Giacconi, S. Murray, H. Gursky, E. Kellogg,
E. Schreier, and H. Tananbaum, Astrophys. J. 178,
281 (1972).

11. S. Grebenev, A. Lutovinov, S. Tsygankov, and
I. Mereminsky, Mon. Not. R. Astron. Soc. 428, 50
(2013).

12. J. B. Hutchings, D. Crampton, and A. P. Cowley,
Astrophys. J., Part 1, 225, 548 (1978).

13. S. A. Ilovaisky, C. Chevalier, C. Motch, M. Pakull,
J. van Paradijs, and J. Lub, Astron. Astrophys. 140,
251 (1984).

14. K. Jahoda, C. B. Markwardt, Y. Radeva, A. H. Rots,
M. J. Stark, J. H. Swank, T. E. Strohmayer, and
W. Zhang, Astrophys. J. Suppl. Ser. 163, 401 (2006).

15. R. I. Kelley, J.G. Jernigan, A. Levine, L. D. Petro, and
S. Rappaport, Astrophys. J. 264, 568 (1983).

16. M.M. Kotze and P. A. Charles, Mon. Not. R. Astron.
Soc. 420, 1575 (2012).

17. H. A. Krimm, S. T. Holland, R. H. D. Corbet,
A. B. Pearlman, P. Romano, J. A. Kennea,
J. S. Bloom, S. D. Barthelmy, et al., Astrophys.
J. Suppl. Ser. 209, 14 (2013).

18. R. Krivonos, M. Revnivtsev, S. Tsygankov,
S. Grebenev, E. Churazov, and R. Sunyaev, Astron.
Astrophys. 519, A107 (2010).

19. F. L. Lang, A. M. Levine, M. Bautz, S. Hauskins,
S. Howe, F. A. Primini, W. H. G. Lewin, W. A. Baity,
et al., Astrophys. J. 246, L21 (1981).

20. A. Levine, S. Rappaport, A. Putney, R. Corbet, and
F. Nagase, Astrophys. J. 381, 101 (1991).

21. A. M. Levine, H. Bradt, W. Cui, J. G. Jernigan,
E. H. Morgan, R. Remillard, R. E. Shirey, and
D. A. Smith, Astrophys. J. Lett. 469, L33 (1996).

22. A. M. Levine, S. A. Rappaport, and G. Zojchenski,
Astrophys. J. 541, 194 (2000).

23. F. Li, S. Rappaport, and A. Epstein, Nature 271, 37
(1978).

24. M. Matsuoka, K. Kawasaki, S. Ueno, H. Tomida,
M. Kohama, M. Suzuki, Y. Adachi, M. Ishikawa, et
al., Publ. Astron. Soc. Jpn. 61, 999 (2009).

25. van derMeer, Ph.D. Thesis (Astron. Inst.Anton Pannekoek,
Univ. Amsterdam, The Netherlands, 2006).

26. D. Moon and S. Eikenberry, Astrophys. J. 549, L225
(2001).

27. W. Pietsch, W. Voges, M. Pakull, and R. Staubert,
Space Sci. Rev. 40, 371 (1985).

28. S. Safi-Harb, H. Ogelman, and K. Dennerl, Astrophys.
J. 456, L37 (1996).

29. N. Sanduleak and A. G. D. Philip, IAU Circ., No.
3023 (1977).

30. W. M. Smart, Celestial Mechanics (Green and Co,
Longmans, 1953).

31. L. Struder, U. Briel, K. Dennerl, R. Hartmann,
E. Kendziorra, N.Meidinger, E. Pfeffermann, C. Reppin,
et al., Astron. Astrophys. 365, L18 (2001).

32. S. S. Tsygankov and A. A. Lutovinov, Astron. Lett.
31, 380 (2005).

33. P. Ubertini, F. Lebrun, G. Di Cocco, A. Bazzano,
A. J. Bird, K. Broenstad, A. Goldwurm, G. La Rosa,
et al., Astron. Astrophys. 411, L131 (2003).

34. N. E.White, Nature 271, 38 (1978).

35. C. Winkler, T. Courvoisier, G. Di Cocco, N. Gehrels,
A. Gimenez, S. Grebenev, W. Hermsen, J. M. Mas-
Hesse, et al., Astron. Astrophys. 411, L1 (2003).

36. J. W. Woo, G. W. Clark, A. M. Levine, R. H. Corbet,
and F. Nagase, Astrophys. J. 467, 811 (1996).

\begin{table*}[t]
\centering
{{\bf Table 3.} Mid-eclipse times for LMC X-4 from the data of various instruments}\label{eclipses}
\vspace{5mm}\begin{tabular}{l|c|l|c} \hline\hline
{Mid-eclipse}& Instrument & {Mid-eclipse}& Instrument \\
      (MJD)  & & (MJD) & \\
\hline
$42829.494(19)^a$   &  SAS-3   & $55385.294(7)$  & INTEGRAL\\
$44956.15(1)^b$   & optics   & $55499.373(6)$  & INTEGRAL\\
$45651.917(15)^c$  & EXOSAT & $55593.729(4)$  & INTEGRAL\\
$45656.154(8)^d$  & EXOSAT & $55595.130(5)$  & INTEGRAL\\
$46447.668(11)^c$  & EXOSAT & $55596.556(13)$  & INTEGRAL\\
$46481.467(3)^c$  & EXOSAT & $55597.938(6)$  & INTEGRAL\\
$47229.3313(4)^{e}$  & GINGA & $55747.234(4)$  & INTEGRAL\\
$47741.9904(2)^{f}$  & GINGA & $55748.645(5)$  & INTEGRAL\\
$48558.8598(8)^e$  & ROSAT & $55751.446(5)$  & INTEGRAL\\
$50315.130(15)^{g,h}$ &  RXTE/PCA & $55938.778(9)$  & INTEGRAL\\
$50740.460(15)^h$   &  RXTE/PCA & $56079.594(6)$  & INTEGRAL\\
$50744.670(15)^h$   &  RXTE/PCA & $56082.424(5)$  & INTEGRAL\\
$51113.680(15)^h$   &  RXTE/PCA & $56109.174(14)$  & INTEGRAL\\
$52647.408(7)^i$  &  INTEGRAL &  $56111.993(6)$  & INTEGRAL\\
$52648.804(6)^i$  &  INTEGRAL & $56119.037(8)$  & INTEGRAL\\
$52892.474(15)^{h,j}$ &  XMM-Newton & $56141.583(8)$  & INTEGRAL\\
$53013.588(4)^i$  & INTEGRAL &  $56142.985(5)$  & INTEGRAL \\
$53016.411(4)^i$  & INTEGRAL & $56292.271(6)$  & INTEGRAL \\
$53172.732(15)^{h,j}$   &  XMM-Newton & $56293.672(5)$  & INTEGRAL\\
$55354.284(9)$  & INTEGRAL & $56295.084(4)$  & INTEGRAL\\
$55355.717(18)$ & INTEGRAL & $56447.192(7)$  & INTEGRAL\\
$55358.531(8)$  & INTEGRAL & $56448.602(6)$  & INTEGRAL\\
$55376.841(8)$  & INTEGRAL & $56450.014(5)$  & INTEGRAL\\
$55378.252(8)$  & INTEGRAL & $56452.824(8)$  & INTEGRAL\\
$55379.656(5)$  & INTEGRAL & $56478.170(4)$  & INTEGRAL\\
$55382.463(5)$  & INTEGRAL & $56483.794(8)$  & INTEGRAL\\
\hline
\multicolumn{4}{l}{}\\ [-3mm]
\multicolumn{4}{l}{$^a$ ``based on the Doppler effect'' (see the text) (Kelley et al. 1983)}\\
\multicolumn{4}{l}{$^b$ from the optical 1976\u20131983 observations (Ilovaisky et al. 1984)}\\
\multicolumn{4}{l}{$^c$ Dennerl (1991)}\\
\multicolumn{4}{l}{$^d$ Pietsch et al. (1985)}\\
\multicolumn{4}{l}{$^e$ ''based on the Doppler effect'' (Woo et al. 1996)}\\
\multicolumn{4}{l}{$^f$ ''based on the Doppler effect'' (Levine et al. 1991)}\\
\multicolumn{4}{l}{$^g$ eclipse was observed partially}\\
\multicolumn{4}{l}{$^h$ a conservative estimate of the error from above (see the text)}\\
\multicolumn{4}{l}{$^i$ see also Falanga et al. (2015)}\\
\multicolumn{4}{l}{$^j$ light curves with eclipses are also given in van derMeer (2006)}\\
\end{tabular}
\end{table*}

\end{document}